\documentclass[fleqn,usenatbib]{mnras}

\usepackage{newtxtext,newtxmath}
\usepackage[T1]{fontenc}

\DeclareRobustCommand{\VAN}[3]{#2}
\let\VANthebibliography\thebibliography
\def\thebibliography{\DeclareRobustCommand{\VAN}[3]{##3}\VANthebibliography}

\usepackage{graphicx}	% Including figure files
\usepackage{amsmath}	% Advanced maths commands

\usepackage{lipsum}
\usepackage[english]{babel}
\usepackage{color}
\usepackage[dvipsnames]{xcolor}

\usepackage{multirow}
\usepackage{cleveref}
\usepackage{romannum}
\crefformat{section}{\S#2#1#3} % see manual of cleveref, section 8.2.1
\crefformat{subsection}{\S#2#1#3}
\crefformat{subsubsection}{\S#2#1#3}
\usepackage{comment}

\title[DESI Machine Learning]{Classifying white dwarfs from multi-object spectroscopy surveys with machine learning}%Machine learning classification of DESI DR1 white dwarfs using spectroscopic and photometric data with UMAP exploitation}

\author[J.\ Munday et al.]{James Munday,$^{1}$\thanks{Email: james.munday98@gmail.com}
Pier-Emmanuel Tremblay,$^{1}$
Ingrid Pelisoli,$^{1}$
Thomas Killestein,$^{1}$
Julia Martikainen,$^{2,3}$
\newauthor
David Jones,$^{2,4}$
Antoine B\'edard,$^{1}$
Paulina Sowicka$^{2,4}$\\
$^{1}$ Department of Physics, Gibbet Hill Road, University of Warwick, Coventry CV4 7AL, United Kingdom\\
$^{2}$ Instituto de Astrofísica de Canarias, E-38205 La Laguna, Tenerife, Spain\\
$^{3}$ Nordic Optical Telescope, Rambla Jos\'e Ana Fern\'andez P\'erez
7, 38711, Bre\~na Baja, Spain\\
$^{4}$ Departamento de Astrofísica, Universidad de La Laguna, E-38206 La Laguna, Tenerife, Spain\\
}

\date{Accepted 2026 February 04. Received 2026 February 04; in original form 2025 December 18}

\pubyear{2025}

\begin{document}
\label{firstpage}
\pagerange{\pageref{firstpage}--\pageref{lastpage}}
\maketitle

% Abstract of the paper
\begin{abstract}
With tens to hundreds of spectra of white dwarfs being taken each night from multi-object spectroscopic surveys, automated spectral classification is essential as part of efficient data processing. In this study, we design a neural network to classify the spectral type of white dwarfs using a combination of spectra from the Dark Energy Spectroscopic Instrument (DESI) data release~1 and imaging from Pan-STARRS photometry. The trained network has a near 100\% accuracy at identifying DA and DB white dwarf spectral types, while having an 85--95\% accuracy for identifying all other primary types, including metal pollution. Distinct spectral or photometric features map into separate structures when performing a Uniform Manifold Approximation and Projection (UMAP) dimensionality reduction. Investigating further and looking at multiple epoch spectra, we performed a separate search for objects that have strongly changing spectral signatures using UMAP, discovering 3 new inhomogeneous surface composition (`double-faced') white dwarfs in the process. We lastly show how machine learning has the potential to separate single white dwarfs from double white dwarf binary star systems in a large dataset, ideal for isolating a single star population. The results from all of these techniques show a compelling use of machine learning to boost efficiency in analysing white dwarfs observed in multi-object spectroscopy surveys, at times replacing the need for human-driven spectral classifications. This demonstrates our techniques as powerful tools for batch population analyses, finding outliers as a form of rare subclass detection, and in conducting multi-epoch spectral analyses.

\end{abstract}

% Select between one and six entries from the list of approved keywords.
% Don't make up new ones.
\begin{keywords}
stars: white dwarfs -- software: machine learning --  binaries: spectroscopic
\end{keywords}

%%%%%%%%%%%%%%%%%%%%%%%%%%%%%%%%%%%%%%%%%%%%%%%%%%

%%%%%%%%%%%%%%%%% BODY OF PAPER %%%%%%%%%%%%%%%%%%

\section{Introduction}
\label{sec:introduction}
In recent years, there has been a rise in multi-object spectroscopic (MOS) surveys \citep{Lamost2012, SDSSmos2013, DESIdesignPaper2016, 4most2019, WEAVEwhitepaper2024}. Their instrumental designs permit hundreds to thousands of spectra to be simultaneously taken in a single telescope pointing, producing spectroscopic data of millions of sources over the course of multi-year surveys. Within each MOS survey consortium, sub-groups focus on specific areas of interest, such as galaxies, dark matter, or mapping local stellar volumes. White dwarfs are a common theme of all wide-scale MOS surveys as objects of great astrophysical importance in galactic evolution as the end-state of the vast majority of stars, being some of the oldest cosmochronometers \citep[][]{Fontaine2001}. There is also a large interest in white dwarfs for MOS survey design as they can be used as flux calibrators, thanks to their relatively smooth continuum and significant space density \citep[][]{Elms2024fluxCalibrator}. Including multiple epoch campaigns targetting time-variable sources, this means that tens to hundreds of WDs are typically observed per night, per MOS instrument \citep[e.g.][]{Manser2024}.

Large data volumes comprising hundreds of thousands of spectra would benefit from efficient classification routines with minimal manual intervention, owing to the computational and human resources required to characterise an exponentially increasing white dwarf dataset. On top of the need to `automatically' separate white dwarfs from other stellar objects through machine learning \citep{2025ApJS..276...53Z,  2025ApJ...988...51P, 2025ApJS..279...36Z}, one desires full categorisation of white dwarf subclasses. To perform in an automated fashion, this requires thousands of test subjects that have been accurately human-classified as an initial training set \citep[e.g.][]{2004ApJ...607..426K, 2006ApJS..167...40E, Gianninas2011, Kleinman2013, Kilic2020sdss100pc, Manser2024, Kilic2025_100pc}, which can then be fed as a model for specific tasks and give general population properties in a matter of seconds. %Even in cases that are not perfectly catalogued, there is the potential to rectify incorrectly labelled spectral types with ever improving data quality, or perhaps even through machine learning, with this perhaps outperforming human capabilities \citep{Vincent2025}, and give general population properties in a matter of seconds. 

The method of spectrally classifying or analysing white dwarfs with machine learning algorithms has recently been explored in the context of white dwarfs polluted by planetary debris \citep{Vincent2023, BadenasAgusti2024, BadenasAgusti2025, Byrne2025}. Their spectra are highly diverse, stemming from differences in the metallic abundances of the accreted material and the metal diffusion timescales of individual white dwarfs. For this reason, their methods need to depend on large, synthetic grids that explore a vast parameter space of white dwarf atmospheric conditions as a training set to detect and constrain metal abundances. Moreover, \citet[][]{GarciaZamora2023, GarciaZamora2025} have used machine learning techniques to show how \textit{Gaia} spectrophotometry can assist in separating white dwarf spectral types, but demonstrate fairly high false-positive percentages (approximately 10--40\% depending on spectral type). This emphasised how machine learning used with \textit{Gaia} spectrophotometry of white dwarfs can hint at spectral types on a very large scale, but that the spectral resolution is much too low to accurately define spectral types, which is a task more suited to higher resolution, MOS survey samples. A comparable performance was obtained by \citet[][]{Vincent2024}, who were able to use machine learning \textit{Gaia} coefficients to obtain precisions of approximately 95\% for the DA (hydrogen Balmer spectral lines) and DB (He~I spectral lines), but struggle to reliably identify any other spectral class.

\citet{Vincent2023} were first to use machine learning to spectrally classify white dwarfs on SDSS's large, MOS survey dataset. Being reduced with a consistent data reduction pipeline, all spectra could be input simultaneously with human-defined spectral classes as a supervised training set \citep{DufourMWDD2017}. Through the full dataset of around 36\,000 spectra that met their signal to noise $>9$ cut, their best neural network model trained on 25\,000 unique targets is able to retrieve a near 100\% accuracy for white dwarfs of spectral type DA, DB and DZ (metal lines, from accreted material or metals in the photosphere). Meanwhile, white dwarfs of spectral type DC (featureless spectrum) and DQ (Swan bands, atmospheric carbon) are found with a 95\% accuracy, DO (He~II spectral lines) at 90\% and DAH (hydrogen spectral lines that are split due to a magnetic field) at approximately 85\%. The work demonstrated an excellent retention of white dwarf spectral type for general population characteristics that rivals the capability of human performance. It is worth noting however that their analysis is limited by the data quality of SDSS, being at a relatively low resolving power of $R=\lambda/\Delta\lambda\approx2000$. With this, their results may lack sensitivity to lower-strength magnetic field DAH objects ($\approx250$--$750$\,kG) that are detectable in DESI but not SDSS, or white dwarfs with a spectrum that includes thin or rare metal lines. \citet[][]{Vincent2025} later expanded on this study by using synthetic spectra as an input, as opposed to the SDSS data products, as an alternative method for machine-driven classification. They were able to obtain an improved detection efficiency to a more diverse set of white dwarf spectral classes using this method.

Since there are around 359\,000 high-confidence white dwarf candidates \citep[][]{NicolaGaia2021} in the \textit{Gaia} data release 3 catalogue \citep[][]{GaiaDR3_2023}, progress in the field for bulk population statistics clearly comes with spectral classification of many more of the unobserved sample of white dwarf candidates and with a larger volume completeness \citep[][]{OBrien2024}. The first data release of the Dark Energy Spectroscopic Instrument \citep[DESI,][]{DESIdr1} delivers approximately 50\,000 co-added spectra of white dwarf candidates based off of just 13 months of data. The increased telescope aperture and spectral resolution compared to previous MOS datasets highlight public DESI data releases as a large and ideal resource to employ with machine learning techniques.

In this proof-of-concept study, we explore the application of a weighted combination of spectroscopic and absolute flux calibrated photometric data to classify spectral classes of candidate white dwarfs found in DESI data release 1 \citep[][]{DESIdr1}. This is the first time that a hybrid approach using spectroscopy and absolute photometry in machine learning classification has been performed with white dwarfs, which is often a vital tool in fitting atmospheric parameters of single- or multiple-star white dwarf systems \citep[e.g.][]{Bedard2017, Munday2024DBL}. Data visualisation tools that are common for machine learning are used to reveal simple and efficient techniques that augment the sample of interesting white dwarf subclasses. Finally, we provide a method of using the DA white dwarfs output from a trained neural network to reveal candidate double white dwarf binary star systems that may masquerade as a single star.

\section{Data handling}
\label{sec:Methods}
\subsection{Spectra}
\label{subsec:methodsSpectra}
We started by downloading all co-added spectra from the DESI data release 1 archive\footnote{\url{https://data.desi.lbl.gov/public/dr1/}}. This dataset includes all spectra previously included in the DESI early data release \citep[][]{DESIeDR2024}, but has since been reprocessed following small reduction pipeline changes. We then performed a crossmatch with the \textit{Gaia} EDR3 white dwarf catalogue of \citet[][]{NicolaGaia2021} using the unique \textit{Gaia} DR2 identifier. To further minimize the number of contaminants in our sample, we only kept objects with a probability of being a white dwarf above 0.5 ($P_\textrm{WD}>0.5$) according to \citet[][]{NicolaGaia2021}. This resulted in a sample of 41\,268 unique sources, each with a single DESI co-added spectrum. Most of these have been specifically targeted as white dwarf candidates in the DESI Milky Way Survey \citep[][]{Cooper2023}, but our selection procedure ensures that white dwarf candidates serendipitously observed in other programs are also included \citep[][]{Manser2024}.

Next, we used the spectral classes labelled in the Montreal White Dwarf Database \citep[MWDD,][accessed 3rd August 2025]{DufourMWDD2017} to assign a spectral class to each white dwarf. The vast majority of unique objects have spectral classes assigned through human inspection, meaning that one can expect for spectral types to occasionally be misentered or misidentified due to human mistakes or from classifying with worse data quality than that presented in DESI. Nonetheless, the MWDD was an excellent resource as a starting point for training a machine learning algorithm. As spoken about in Section~\ref{subsec:ReclassificationOfSpectralType}, we later iteratively corrected the spectral types of incorrect entries to the best of our abilities\footnote{Within the MWDD, all spectral classifications from the DESI early data release spectra presented in \citet[][]{Manser2024} are included.}.

We then crossmatched the white dwarfs in DESI with all unique target entries in the MWDD and were left with 19\,516 targets. A soft cut on the data quality of the spectra was performed to remove any data quality issues as indicated by non-zero bitmasks given in the data products or when the inverse variance for a significant proportion of the spectrum was zero. We also remove cases where the continuum signal-to-noise ratio in the blue arm has a median value less than 2, deemed to be a misplaced fibre or an otherwise useless spectrum for our analysis. This left 19\,292 unique white dwarfs.

There is a very large diversity in the spectral classification scheme listed on MWDD, and, for the sake of having a large training sample for each class, we limited to objects that were of spectral type DA, DAH, DAZ/DZA, DB, DBZ/DZB, DC, DO/DOA/DAO, DQ, DZ or cataclysmic variables (CV). Upon experimenting with neural network architectures and class weightings in Section~\ref{subsubsec:ModelPerformance}, it proved too challenging to reliably separate DAB/DBA white dwarfs from DAs and DBs because of their relatively low number of sources and lack of unique spectral features, so they were removed from the training sample.  Other classes had 1--20 total occurrences in the database, %with the exception of DAO/DOA (which we wrap into the DO category), 
so are also not a reliable class to validate our model owing to the small-number statistics. Any instance where the spectral class is uncertain in the MWDD (denoted by a single colon at the end of the spectral type) was ignored, as well as cases of a white dwarf with a fainter main sequence companion. Selecting just these spectral types (after reclassification of some, see Section~\ref{subsec:ReclassificationOfSpectralType}), spectra of 17\,614 unique entries remained.

All DESI spectra are sampled in 0.8\AA~intervals from 3600--9824\AA, and just the data in blue arm (3600--5800\AA) and red arm (5760--7620\AA) were utilised in this study, since spectra from the near-infrared arm are not particularly useful for white dwarfs as few spectral lines appear at these wavelengths. A machine learning algorithm needs all input data on the same scale, so we normalised all spectra but processed data from the blue and red arms separately. We first isolated all data between 3720--7500\AA, removing the edges because of a worse data quality. Since we wanted the same normalisation to all spectra in the DESI dataset, we masked out regions where spectral features are commonly found in white dwarfs for continuum normalisation (Appendix~\ref{Appendix:masking}). For the blue arm data, we first binned the data in 40\AA~bin widths, applying a 3$\sigma$ clipping of points within each bin. A 7th order polynomial was fit to the binned data points and $+2.5\sigma$ or $-3\sigma$ sigma outliers\footnote{The sigma clipping threshold was chosen to be higher for points below the continuum because any artificial drop mixes with absorption features, which should not be removed. For points above the continuum, the only real features are emission lines in CVs, which occasionally had some data points clipped for very strong emission lines, but this did not noticeably impact any automated classification since the emission signature was preserved and still easily recognisable.} in the residuals was removed for all points besides those below 3900\AA~(the reason being to maintain a good normalisation for objects that exhibit Ca~II absorption lines). Data between 5500--5800\AA~was then removed, since the edges of spectra in polynomial fits can introduce strong artefacts. We binned the red arm co-added spectrum in widths of 24\AA~with a 2.5$\sigma$ clipping in each bin, fit a 4th-order polynomial and removed $+2.5\sigma$ or $-3\sigma$ outliers in the residuals for all wavelength bins. This process was repeated 3 times with any outliers in the residuals of the binned polynomial fit clipped on each iteration. Since few spectral lines exist in far red wavelengths, we removed all data above 6812\AA~to ensure that the network focused on the spectral lines of interest. The normalised blue arm and red arm data were finally combined and $5\sigma$ outliers in the resultant normalised spectrum were clipped, with flux measurements for the clipped data points replaced by a linear interpolation of the two nearest neighbours.

After analysing all data, clear outliers appeared that are either detector or DESI reduction pipeline related. This most notably occurred in the wavelength range 4285--4410\AA~where the absolute flux calibration is systematically lower than the neighbouring spectral regions \citep[][]{Manser2024}. This defect coincides with an occasional drop in signal-to-noise and, since this wavelength range includes absorption lines of interest such as H$\gamma$ (4340\AA), we decided to mask the entire wavelength range within which the extended wings of H$\gamma$ may be present. Doing so creates an artificial gap in wavelength coverage for the input data but bears no meaning to an untrained neural network, so its impact to the performance results is small and DESI DR1 specific.

%BLUE ARM DATA <6000A:
%bin the points with width of 50 points  (0.8*50 = 40A), with 3 sigma clipping

%coeffs = np.polyfit(binned_wavelength, binned_flux, deg=7)#, w=1/binned_flux_err)

%detect +3 sigma -2.5 sigma outliers, in the fit to the binned curve

%mask <5500A for blue side

%RED ARM >4950A:
%bin the points with width of 30 points  (0.8*30 = 24A),  with 2.5 sigma clipping

%coeffs = np.polyfit(binned_wavelength, binned_flux, deg=4, w=1/binned_flux_err)

%detect +3 sigma -2.5 sigma outliers, in the fit to the binned curve

%loop 3 times

%mask >5500A for red side

\subsection{Photometry}
\label{subsec:Photometry}
The fact that all DESI observations occur above a declination of approximately $-25^\circ$ means that all-sky and northern photometric surveys include flux-calibrated photometry for almost every target of interest, which can be placed on an absolute flux scale using a distance obtained from the inverse of the \textit{Gaia} parallax. The most important aspect here is to maximise the completeness of photometric points for our sample of white dwarfs. We trialled broadband photometry from \textit{Gaia}, SDSS \citep{SDSSdr16}, Pan-STARRS \citep[][]{Panstarrs} and all combinations of each to judge their advantage in improving the accuracy of spectral classification. Overall, and like in the case of previous work \citep[][]{Munday2024DBL}, we found that \textit{Gaia} photometry provided no noticeable improvement compared to SDSS or Pan-STARRS because of its much wider transmission curve. Wider spectral coverage using other photometric surveys, such as the Galaxy Evolution Explorer \citep[GALEX,][]{GalexMission2005} or the Two Micron All Sky Survey \citep[2MASS,][]{2massMission2006}, would be ideal to extend spectral coverage. This is because the flux in the ultra-violet data helps to separate a white dwarf's atmospheric opacity in spectral classification \citep[e.g.][]{Blouin2023b, Kilic2025mergerRemnantsFUV} and 2MASS would be better for near infra-red flux excesses from a stellar companion \citep[e.g.][]{Wachter2003, Hoard2007}. That said, the low photometric precision for fainter objects in 2MASS and the lack of sky completeness in GALEX makes their inclusion disadvantageous for the task.
%and we consider an extended spectral energy distribution wavelength range as an improvement to our methods in future years with new telescope surveys.

We also investigated the usage of \textit{Gaia}~XP spectra or coefficients, but again found no visible benefit since, across the full sample, the poor quality in the bluer end of the data adds confusion noise. However, we did find that synthetic SDSS and Pan-STARRS photometry generated from the \textit{Gaia}~XP coefficients in GaiaXPy \citep[][]{GaiaXPdaniela_ruz_mieres_2024_11617977} gave flux measurements of comparable quality in all but SDSS $u$. Any $u$-band data from SDSS or generated from the \textit{Gaia}~XP coefficients was suitable for some objects when using an error below 0.3\,mag, but the reduction in the training set sample size proved to negatively impact the neural network efficiency, outweighing the positive of a wider photometric coverage.  The best compatibility with the white dwarf sample appeared to come with using Pan-STARRS photometry in the $g, r, i, z$, and $y$ bands and to insert SDSS $g, r, i, z$ photometry if there are null entries for any object or filter. Use of Pan-STARRS survey data was prioritised. At times, the Pan-STARRS $y$ data is the only null or high error entry, likely due to a worse sensitivity to dimmer objects in the near-infrared for ground based observations compared to that of the optical, or increased flux contamination from nearby objects. In these cases, we insert \textit{Gaia}~XP generated synthetic photometry in the $y$ filter if it has an error less than 0.3\,mag.

Of the 17\,614 spectra, the steps of restricting to targets with the desired spectral types listed in Section~\ref{subsec:methodsSpectra}, restricting to sources with photometry for the $g, r, i, z, y$ bands and removing unclear cases from the training sample (Section~\ref{subsec:ReclassificationOfSpectralType}) provided a reduction down to 16\,039 unique targets for the purpose of training/testing a neural network. All spectra and photometry for each object were finally dereddened using the mean extinction values quoted in the catalogue of \citet[][]{NicolaGaia2021} and $E(B-V)=A_V/3.1$.

\subsection{Reclassification of spectral type}
\label{subsec:ReclassificationOfSpectralType}
While we used the MWDD as an input catalogue, there are inaccuracies that are rectifiable with new and in some cases improved data at hand. Iteratively training a neural network and looking at false negatives or low-probability solutions helped to flag discrepancies that were inspected visually. Reclassifying spectral types proved especially relevant for the DAH class where many were classified as DA white dwarfs in the literature, stemming from the increased spectral resolution of DESI data compared to lower-resolution data that has often been used for white dwarf spectral classification in custom setups or SDSS. 

There are also cases where there is a lack of clarity in the DESI data for if an object is a DA, DAZ/DZA or DAH, perhaps because of an inadequate signal-to-noise ratio to be certain of a real feature. Neither a human nor a machine would be able to confidently assign a spectral class in these situations, so they were removed completely from the training set to improve purity of the input catalogue, as was the same for borderline DA/DC, DB/DC or DC/DQ cases. This means that borderline cases will result in low maximum probabilities, as should be the case instead of being erroneously placed in one specific class. 

35 DB white dwarfs have been reclassified as a DBZ/DZB white dwarf, which largely comes from the clear presence of a Ca~K absorption line. The full list of objects and spectral classes that are used in our training sample is uploaded at \textsc{https://github.com/JamesMunday98/MachineLearningWD}.

\section{Neural network}
\label{sec:neuralNetwork}
A neural network was built using the TensorFlow \citep[][]{tensorflow2015-whitepaper} and Keras \citep[][]{chollet2015keras} modules in Python. The architecture of the system starts with feature extraction from three, 32 neuron temporal filters of length 3 pixels (2.4\AA), 11 pixels (8.8\AA) and 23 pixels (18.4\AA) to identify narrow, intermediate and wide patterns in the spectra. Two layers follow of length 64 and 128 neurons and width 15 (12\AA) and 23 pixels (18.4\AA) to look for patterns between the three feature extractions. Then finally, to transform these patterns into a predicted spectral class, a shared base was instated with three fully connected layers with 256, 128 and 64 neurons, respectively. Each dense layer employed a rectified linear unit \citep[][]{householder1941theory, nair2010rectified} activation function and L2 weight regularisation of $\lambda=10^{-4}$, which served to mitigate overfitting by adding a penalty term to the loss function that is proportional to the sum of the square of the weights. After each dense layer, batch normalisation \citep[][]{2015arXiv150203167I} was applied to stabilise training and accelerate convergence, followed by dropout layers \citep[][]{JMLR:v15:srivastava14a} with dropout rates of 0.4, 0.3, and 0.2, respectively, to further improve how well a trained model performs on new, unseen data.

Two independent output heads were defined. The first was to find the primary atmospheric class, while the second head was to search for metal lines and only shares neurons for learning between the DA, DB and DZ classes; all other primary atmospheric classes were ignored from the loss function. Therefore, this head was responsible for characterising between DA versus DAZ/DZA or DB versus DBZ/DZB, and the DZ class being linked helped to improve learning rates. The model was trained using the Adam optimiser \citep[][]{adam2014method} inside the Keras module with an initial learning rate of $3\times10^{-4}$.

The vast majority of the unique targets are of spectral type DA. Attempts to address this class imbalance by giving every class an equal weight proved less effective at accurately characterising a test sample than applying no weighting at all. This occurred because the DAZ/DZA class suffered from a model that focuses less on the DA objects, consequential of our approach in obtaining the primary atmospheric class followed by the search for the metal component. We found best results when the weighting of every DA was set to one-quarter of that for all other targets, while the weighting of DAZ/DZA was weighted three times higher. Even with this included, the DA class still had an approximately 50\% higher weighting than all other spectral classes combined, but we viewed a model that is slightly more sensitive towards DA white dwarfs to be advantageous so that a clearer separation is made between the DAH or DAZ/DZA class, which can be marginally detected for cases with low magnetic field strengths or thin metal lines, respectively. Lastly, due to their relatively small number in the training set and variety in emission line strengths, we weighted the CV class to be three times stronger.

The final weighting to consider is that assigned to the DESI spectral component and the Pan-STARRS photometric component. The few thousand DESI spectral points outnumber the five of Pan-STARRS, hence, we trialed relative weightings of the two through scaling of the variance of the two categories. We found that a 30\% photometry to 70\% spectroscopy weighting is optimal for maintaining a high accuracy across all spectral types and we employed this in our final model. Other weightings of photometry between 25--50\% gave extremely similar results, so we intentionally focus the network to have a better sensitivity to unique spectral features and use the photometric component to assist single star classification, but not dominate. No flux errors are handled by the neural network to avoid under-weighting of spectral signatures in certain spectral windows for the less populated classes. This means that each spectroscopic point was weighted identically to all others, and the same for the photometric data points.

%\section{Results}
%\label{sec:Results}
\section{Single star classification}
\label{sec:SingleStar}
\subsection{Model performance}
\label{subsubsec:ModelPerformance}

\begin{figure*}
    \centering
    \includegraphics[width=\textwidth,clip,trim={0.15cm 0.35cm 0.25cm 0.25cm}]{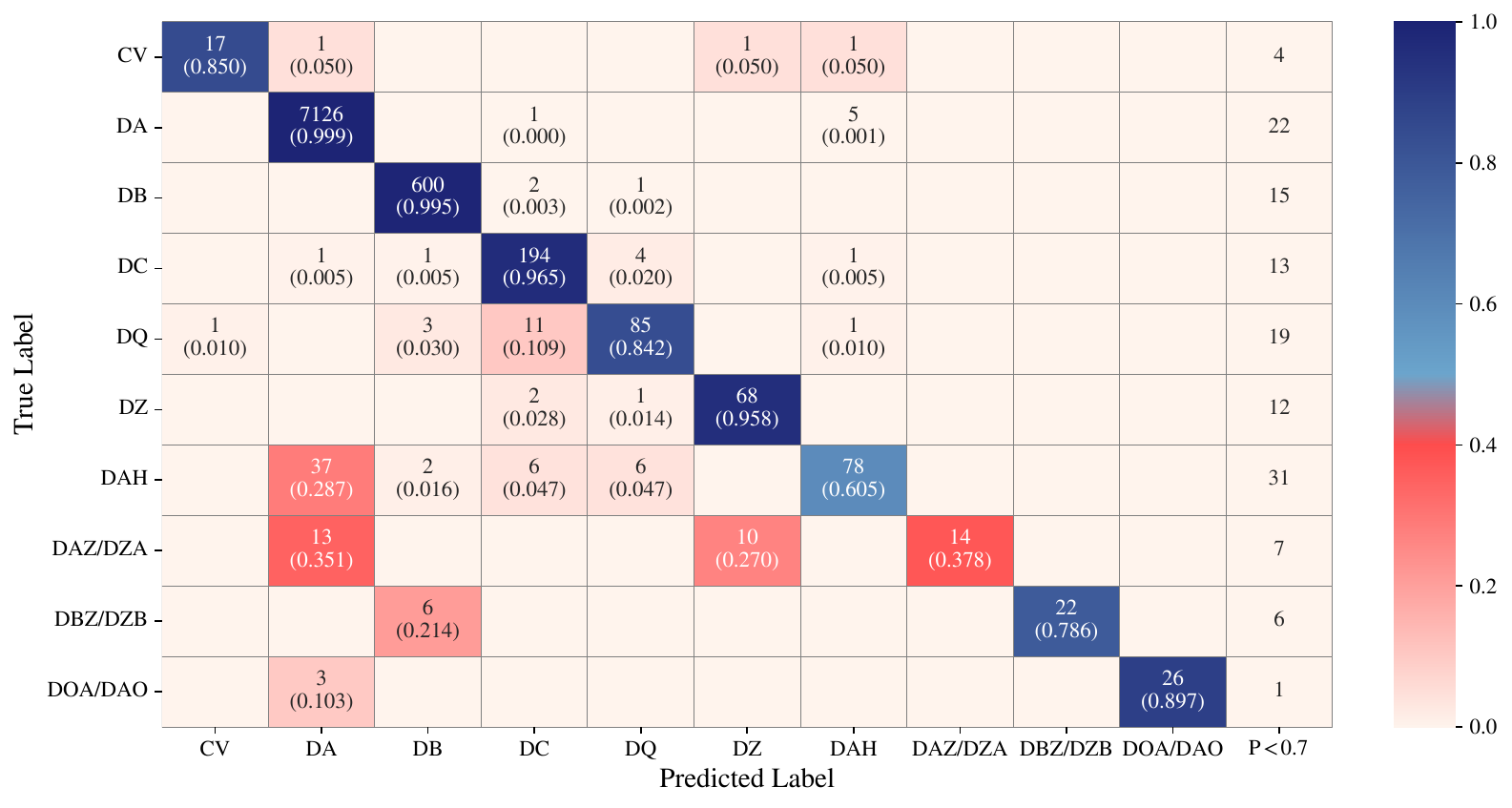}
    \caption{A 5-fold confusion matrix, with results from each individual 80\%--20\% training--test data split combined into one confusion matrix. Non-empty cells state the number of systems on top of the fraction of objects that fall into the category for each true label. The final column shows the number of systems for which the predicted label does not surpass a 70\% confidence level, and these systems are removed from the performance statistics of the model. These results show the approximate performance we can expect of our final model on a new, unseen dataset. The final best-fit model is trained on 100\% of the data, hence these numbers can be assumed as a minimum classification accuracy.}
    \label{fig:ConfusionMatrices}
%\end{figure*}

%\begin{figure*}
    \centering
    \includegraphics[width=\textwidth,trim={0.0cm 0.3cm 0.0cm 0.25cm}]{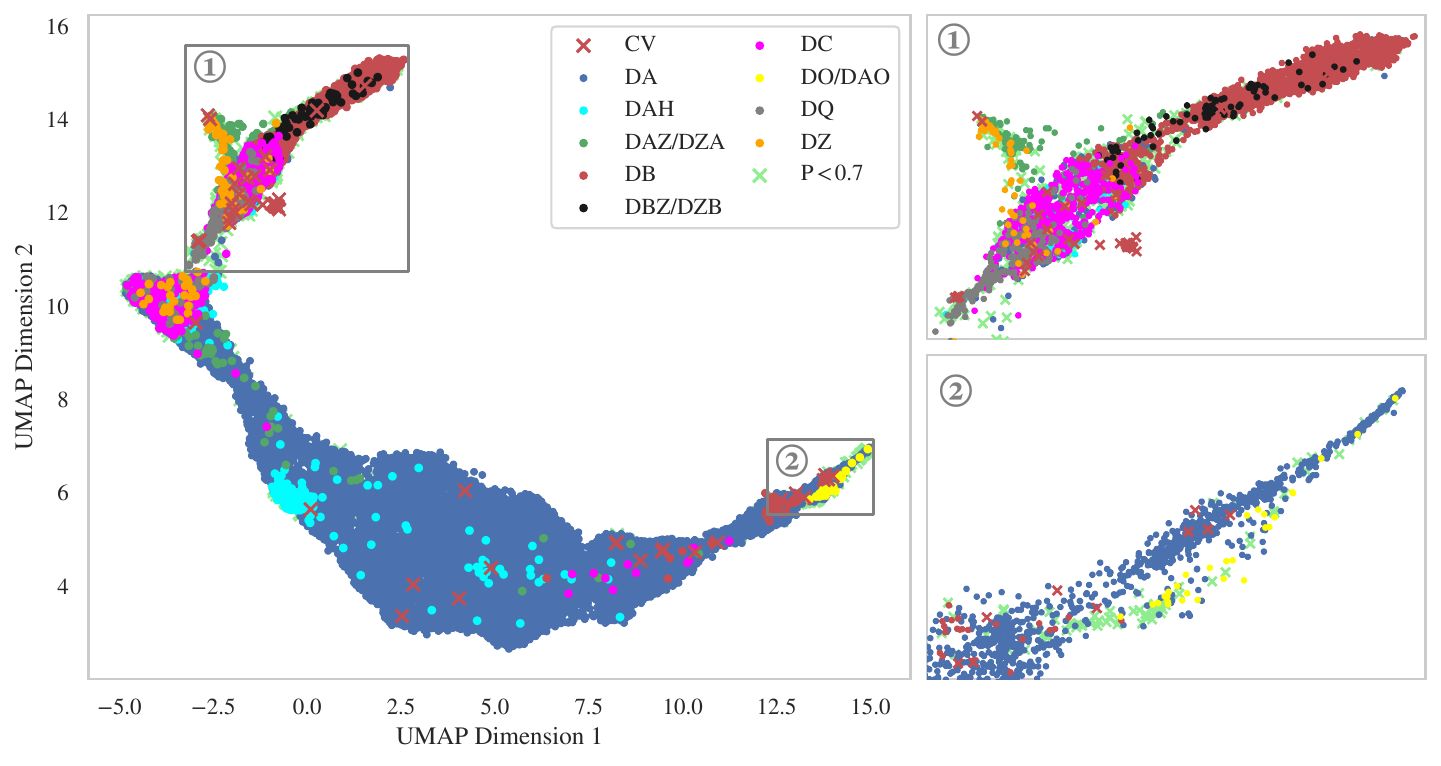}
    \caption{A projection of all co-added spectra in the DESI DR1 dataset that meet our quality cuts (21\,344 unique targets) onto 2-dimensional UMAP space. Points are colour coded by the stellar subclass predicted by our best-fitting neural network. The curvature seen in the figure in the DA primary class (from the far left curved around to the far right) is representative of the colour and absolute magnitudes of the source, where stars with higher effective temperatures lie towards the right. Similarly, non-Balmer line sources are grouped at the top left (besides DO/DOA/DAO white dwarfs, see Section~\ref{subsubsec:DataVisualisation}). Two zoomed insets are displayed around the non-DA (1) and the DA or DO/DOA/DAO (2) primary classes. Evidently from the purity of systems in certain groupings, UMAP offers an immediate means to classify DZ/DZA white dwarfs, DB/DZB white dwarfs, DQ white dwarfs, DAH white dwarfs or CVs with just the UMAP coordinates. Some CVs are found away from the dense grouping near coordinates (0,12), likely due to data reduction issues or any emission lines present.}
    \label{fig:UMAPsingleClassifier}
\end{figure*}

All results presented now have a final signal-to-noise cut to ensure that the median value of the continuum (the non-masked region described in Appendix~\ref{Appendix:masking}) between the wavelength range 4000--5000\AA~is greater than 5 per resampled pixel, and the same for the median value between 3780--4000\AA. One could likely probe slightly lower for primary class classifications, but we noticed that the recovery of magnetic and metal polluted objects deteriorates when further lowering the signal-to-noise threshold. The pixel sampling of DESI spectra is about 1.5$\times$ finer than that of SDSS, and in SDSS a signal-to-noise greater than 8 was needed for accurate human classification of white dwarf spectral type \citep[][]{Kepler2013, Kleinman2013}, so the signal-to-noise threshold per Angstrom is similar. As a result, the final sample size of objects that pass all cuts and have an entry in the MWDD amounted to 8\,481 unique objects.

Fig.~\ref{fig:ConfusionMatrices} shows a confusion matrix of best neural network when trained on the entire final sample. Naturally, a model when tested on itself should perform better than when tested to unseen data, so we split 80\% of the data in each class being for model training and 20\% in each class being for testing (which is unseen by the machine). We performed a five-fold cross-validation so that all spectra were tested, and the performance metrics from each iteration were combined to generate a confusion matrix. This should give a good impression of the minimum performance one could expect from a 100\% training data model when applied to an unseen dataset. Subclasses are confidently defined when the probability of a class exceeds 0.7 or is assigned the $P<0.7$ class otherwise.

Our best model performs extremely well, with the accuracy of identifying DA, DB, DC and DZ objects being near 100\%. The CV class performs very well at identifying strong emission lines, while incorrectly predicted outliers have weak emission in each case -- here, emission is seen at H$\alpha$ with small full-width at half maximum values and never exceeds the continuum flux level, leading to confusion. The DAH class has 61\% accuracy with some confusion with the DA class, primarily originating from cases of deep Balmer lines with low magnetic field strengths. In addition, higher magnetic field strengths at times get confused with featureless DC spectra or hot DQ white dwarfs, having a notable fraction of mis-predicted cases. The DAZ/DZA class is always identified as being either a DA, DAZ or DZ white dwarf. The DA confusion occurs either when a Ca~II~H and~K doublet is very narrow ($\lesssim5$\AA) or when there are metals that are under-represented in the training set. For instance, there are two cases where the only visible metal lines are the Na~I resonance doublet at 5889.95\AA~and 5895.92\AA. For the cases where DAZ/DZA objects are confused for a DZ, this occurs when the Ca~II lines are very deep while a detection of H$\alpha$ is small, and this could also be a strong point of improvement to the model with a larger sample of DAZ/DZA white dwarfs in the future. A similar case is seen for the DBZ/DZB class where a few objects are confused as a DB white dwarf, because the metal detection is faint compared to the various He~I lines present. Cases where the DQ true labels have a predicted probability less than 0.7 or when they are confused with a DC stem from a faint detection of carbon Swan bands, but real and faint detections were visually identifiable when smoothing non-normalised spectra. Likewise, even though the accuracy is very high for the DC class, the $P<0.7$ cases originate from a DC that could be confused with a DQ because of lower signal-to-noise ratios.

\citet[][]{Vincent2025} trained their neural network on private, synthetic spectra. Although they show a clear improvement in the spectral classification accuracy compared to \citet[][]{Vincent2023}, this makes it difficult to perform direct comparison to our results. \citet[][]{Vincent2023} however trained their network based on the reduced SDSS data products in the same light as we do with DESI. They tailor their network to be sensitive towards a slightly different set of classes. Nonetheless, there are some spectral type crossovers. They too are able to retain a near 100\% accuracy of finding DA and DB white dwarfs, and obtain a similar accuracy for DC white dwarfs of 96\%. %The likely reason for a higher accuracy in our model is because of the hybrid approach in using DESI spectra and Pan-STARRS photometry for all targets, whereas \citet[][]{Vincent2023} use normalised spectra alone for spectral classification and \textit{Gaia} photometry in an earlier step to flag white dwarf + main sequence star candidates. 
Though their model does search for DZ objects, they chose not to separate into further sub-categories, and as such we do not comment on differences between model performance. \citet[][]{Vincent2023} outperform our model in the CV category, which is a consequence of our best model having a small training size of 12 CVs, compared to 221 CVs in theirs. Our model displays a slightly lower detection of DAH white dwarfs (60.5\% versus their 83.5\%), but this is likely because of the improved sensitivity that DESI has in detecting lower magnetic field strengths and many DA white dwarfs in the SDSS sample should actually be listed as a DAH (Section~\ref{subsec:ReclassificationOfSpectralType}). Thus, with the new and improved DESI data at hand and the corrected spectral classifications, their DAH accuracy should be lower. In both our and their models, the false negatives mostly appear with DAHs classified as a DA. %DO white dwarfs are searched for in both classifiers, but we chose to remove DAO/DOA white dwarfs from the training set while \citet[][]{Vincent2023} do not, so results should not be directly compared and we would likely retain a similar confusion rate between DOs and DAs like they witness.

It is worth a special mention that \citet[][]{Vincent2023} decided to separate DQ and hot DQ objects. Their spectral signatures can differ significantly, but needing to separate DQs into two categories is circumvented by the use of flux-calibrated Pan-STARRS photometry, while the small number of hot DQ objects justifies combination as well. Also worth mentioning is that one should note the much larger sample size of spectrally classified white dwarfs that SDSS currently has over DESI, meaning that the smaller number statistics presented for some categories could be subject to at least a couple of percent changes with future data. A larger catalogue of DESI spectral types would increase the training data set, our neural network would improve in sensitivity significantly for the less-highly sampled classes, and we would hence expect to obtain higher performance metrics.

\subsection{Data Visualisation}
\label{subsubsec:DataVisualisation}
The output of passing a normalised spectrum and Pan-STARRS photometry to the neural network is the probability of an object belonging to each class. To assist in interpretation of the results in a more human-friendly way, we constructed a Uniform Manifold Approximation and Projection \citep[UMAP,][]{McInnes2018umap} embedding of the photometry plus spectroscopy that is input to the network for non-linear dimensionality reduction. In this algorithm, local relationships within the dataset in high dimensional space are abstracted by a lower dimensional representation, attempting to capture and preserve local and global structures as well as possible. Multiple data visualisation algorithms could be used to identify white dwarf class substructure in a large sample \citep[such as t-SNE,][]{Byrne2024}, but we chose to focus on UMAP due to its strong ability to perform well on large datasets \citep[e.g.][]{Kao2024} and because one can compare two different input samples on the same UMAP coordinate space, which is not possible for some other techniques.

Just like the Hertzsprung-Russell diagram has been notoriously used for analysis of stellar objects, exploration of UMAP space is a powerful tool to immediately identify white dwarf spectral types. From a projection with both the training and the rest of the DESI DR1 spectra, we see in Fig.~\ref{fig:UMAPsingleClassifier} a population of DAH white dwarfs that have clearly Zeeman split Balmer absorption lines which fall to the side of the grouping of DA objects. Higher magnetic field objects are scattered because of their decentralisation from H$\alpha$. DZs and some DAZ/DZA objects together form an island away from DC white dwarfs, as is the case for CVs. Many DQ white dwarfs with strong Swan bands fall just above the point where some DA and DC white dwarfs join together in UMAP coordinate space. Moreover, the first UMAP dimension shows some correlation with the absolute brightness of the sources, causing brighter (hotter) white dwarfs to be located towards the right of the plot and cooler white dwarfs towards the left. Many DO/DOA/DAO white dwarfs are in a separate strip to hot DA white dwarfs because of hotter temperatures but unique He~II line signatures. Other substructures in UMAP coordinate space can be exploited as an effective tool in the context of identifying interesting spectral features, for instance the Balmer line emission profile of DAHe white dwarfs \citep[][]{Greenstein1985DAHe, Reding2020DAHe, Gansicke2020DAHe, Reding2023DAHE, Manser2023DAHe}.

%\begin{figure}
%    \centering
%    \includegraphics[width=\columnwidth, trim={0.3cm 0.4cm 0.2cm 0.3cm}]{DAHe.pdf}
%    \caption{The same UMAP construction as Fig.~\ref{fig:UMAPsingleClassifier} with DAHe white dwarfs plotted in black circles over any other object in grey. All points group in a confined location, where isolating nearby points in UMAP coordinate space could drastically decrease the time needed for new DAHe discovery. The same technique can be applied to accelerate the discovery of other white dwarf spectral types based on the interest of a study.}
%    \label{fig:DAHeUMAP}
%\end{figure}

%Substructure in UMAP coordinate space can be exploited as an effective tool in the context of identifying other interesting spectral types. %To demonstrate the power of this and, because of its strong applicability to DESI, we take the list of DAHe white dwarfs as identified in \citet[][]{Manser2023DAHe} on top of other DAHe objects in the literature \citep[][]{Greenstein1985DAHe, Reding2020DAHe, Gansicke2020DAHe, Reding2023DAHE}. Their UMAP coordinates are plotted in Fig.~\ref{fig:DAHeUMAP}, congregated in the connecting region between the DA and non-DA objects. \citet[][]{Manser2024} identify that DAHe object all cluster in a common area in the \textit{Gaia} Hertzsprung-Russell diagram. In a similar light, one can narrow down their congregation to a finer region through using the higher dimensionality reduction UMAP technique with the absolute photometric magnitudes combined with the DESI spectra.

The graphical representation of UMAP parameter space is unique to any new dataset, as would be the case with a UMAP analysis of all spectra found in future DESI or other MOS survey data releases, but the tools outlined could trivially be expanded as an excellent way to augment under-represented subclasses of white dwarfs. One just needs to compile a catalogue and search for patterns in UMAP space, which could be to find new DAHe objects, merger remnants \citep[][]{Kilic2024DAQ}, diverse types of accreting binaries \citep[][]{Ranaivomanana2025}, specific metals in polluted white dwarfs \citep[][]{Hollands2021, Kaiser2021}, and much more. Additionally, projection of new data in the coming years onto the UMAP dimensions presented in Fig.~\ref{fig:UMAPsingleClassifier} could be performed to rapidly identify clear substructures.

\subsection{Validating the usage of the neural network for rarer white dwarf spectral types}
\label{subsubsection:ValidatingUsageOfNeuralNetwork}
Obviously, we can not train a neural network to be equally sensitive to every white dwarf spectral type ever identified as some spectral classes have only 1 or 2 known systems \citep[e.g. DAe white dwarfs, ][]{Elms2023} and/or are too similar to other classes in the network (e.g. DBAZ white dwarfs). Commonalities can be searched for, like performed for the metal polluted cases in our methods (Section~\ref{sec:neuralNetwork}), but at times the spectral signature of, for example, the fourth characterising letter in a white dwarf spectral type has too slim of a spectral signature to decisively separate it from a two or three letter classification. This worsens the applicability of our neural network to large, diverse datasets, and we now address these pollutants by segmenting with UMAP coordinates.

Since DA white dwarfs have the largest weight in the training sample, and hence there is a slight bias towards the DA subclass, one could expect that an `unseen' class with any hydrogen absorption features will be classified as a DA. In the UMAP space of Fig.~\ref{fig:UMAPsingleClassifier}, most DA white dwarfs cluster separately to all other non-DO/DOA/DAO spectral types. So, if we take all objects that fall into the zoomed inset showing non-DA and non-DO/DOA/DAO sources in the trained sample (which are correctly identified objects) and search for how many DA white dwarfs fall in the coordinate space when transformed onto the same UMAP projection, we can compare the fractional pollution of the training sample to the fractional pollution of the full DESI sample.

When selecting the region in Fig.~\ref{fig:UMAPsingleClassifier} where most DA white dwarfs lie (UMAP Dimension 2 < 11), we find that 95.3\% of the targets have a human-classified DA spectral class, while the machine-classified spectral class for the full DESI sample returns 93.4\%. These fractions are very similar. When performing the same test to search objects without hydrogen absorption lines that are classed as DA white dwarfs (UMAP Dimension 2 > 11), the human classification returns 5.5\% of objects as DA white dwarfs while the machine classifies 8.4\%. Visually inspecting these cases in search for an explanation, we saw that DAB/DBA white dwarfs are classified as a DA (but typically with a 0.1--0.3 DB probability) and noticed how some main sequence star plus white dwarfs cases lie in this UMAP space. The main sequence stars are undetected in the spectral range selected and in all cases the dominant signature of a DA white dwarf prevails, causing them to be classed as DA white dwarfs even with a $z$- or $y$-band excess. Hence, `unseen' classes with any hydrogen absorption features do indeed get preferentially assigned to be a DA. There were no cases where, for example, an obviously DB/DC was classified as a DA in this region.

The overall correctness of finding DA white dwarfs is relatively consistent between the human and machine approach. However, we exaggerate how deviations from the trained classes will not always be reliably detected in an automated fashion even if this could seem obvious to the eye of a human expert. One should hence be cautious on the applicability of any machine learning model to the purity of the objects for the analysis they wish to perform. Automated machine learning techniques offer a fast and effective way to manage or group large spectroscopic samples, but, without a large training sample for every white dwarf subclass, they will struggle to be as precise as human vetting \citep[][Swan et al. in prep]{Manser2024}. It is true that machine learning algorithms have great potential in correcting spectral types in the literature \citep[as postulated in][]{Vincent2023}, but in the vast majority of times this comes with a new analysis of improved data quality, like performed in Section~\ref{subsec:ReclassificationOfSpectralType}, instead of an advantage over human abilities.

\subsection{Applying the trained neural network to all white dwarfs in DESI DR1}
Now verifying that our trained neural network model is suitable for wide-spread usage, we applied it to all white dwarfs in DESI DR1 in the same way as was performed in Section~\ref{sec:Methods}. The starting 41\,268 $P_\textrm{WD}>0.5$ objects were reduced to 21\,344 objects that pass our signal-to-noise cuts and have compatible photometric data. A total of 17\,041 objects have the DA (79.84\%) classification, 373 DAH (1.75\%), 226 DAZ/DZA (1.06\%), 1510 DB (7.07\%), 76 DBZ/DZB (0.36\%), 727 DC (3.41\%), 29 DO/DOA/DAO (0.14\%), 226 DQ (1.06\%), 124 DZ (0.58\%), 67 CV (0.31\%) and 945 have $P<0.7$ (4.43\%). The $P<0.7$ objects largely fall in areas of the UMAP space in Fig.~\ref{fig:UMAPsingleClassifier} where two subclasses meet, for instance DA and DAH, DA and DC, DAZ/DZA and DZ. $P<0.7$ could also stem from confusion with the noise on individual pixels or data reduction artefacts. For the same reason of DESI data reduction artefacts, where there are pixels above the continuum that mimic the appearance of emission lines, the network misclassifies some spectra as CVs, recognisable as points that lie far away from the main CV grouping near coordinates (0,12) in Fig.~\ref{fig:UMAPsingleClassifier}. Finally, although an attempt was made to remove non-white dwarf objects by selecting $P_{\rm WD}>0.5$ objects from the catalogue of \citet[][]{NicolaGaia2021}, non-white dwarf objects will likely be present in the 21\,344 objects sample. Those that do not photometrically or spectroscopically resemble one of the white dwarf training classes are not familiar to the machine learning classifier, hence they can not be confidently classified and fall into the $P<0.7$ category.

\section{Searching for spectral type changes}
\label{subsec:resultsSpectralTypeChanges}
All analysis in this study for single star classification utilised co-added spectra, but individual exposures have the power to reveal variability in a whole host of ways. Depending on the time of observation, some rare white dwarfs can be observed as a DA, DB or a mix of the two. This was reported, for example, by \citet[][]{Caiazzo2023} who discovered these spectroscopic variations in ZTF~J203349.8+322901.1, concluding that the presence of a close companion cannot explain the sudden spectral type switches. Instead, these author conclude that the object has two faces that come into view over a 14.97\,min rotational period, with the two faces differing in composition. \citet[][]{Bedard2025} provide an explanation that a surface hydrogen layer of varying thickness can naturally give rise to the time-dependent changes in the observations of ZTF~J203349.8+322901.1. The same effect was witnessed by \citet[][]{Pereira2005} in GD~323, and the inhomogeneous surface composition hypothesis was proposed as an explanation there as well, but for a case where the spectral changes are less stark. \citet[][]{Moss2024, Moss2025} discovered two such systems, establishing these objects as a class. The full selection of these white dwarfs include both magnetic objects and white dwarfs with undetected magnetic fields \citep[][]{Moss2025}, and interestingly they span a wide range of atmospheric parameters and rotational periods.

Since the observational changes can be so jarring, we found the outlined visualisation methods to be a perfect application to search for more objects of this nature. Important to note is that an approach that solely relies on our machine learning classifier (Section~\ref{sec:SingleStar}) is less sensitive to small changes in spectral features because the machine will classify a white dwarf spectral type by the dominant spectral features present. For the situation above, a faint detection of hydrogen may be a noticeable change in spectroscopic signature, but a dominant helium line signature drives the machine to classify the object as a DB white dwarf. The utilisation of visualisation techniques circumvents the need to train a machine on under-populated, branching spectral types (in this case, the small number of DAB/DBA targets would branch between DA and DB), making it better suited to the task.

We first downloaded all exposures for each white dwarf in the full \citet[][]{NicolaGaia2021} catalogue with any probability of being a white dwarf, totalling 49\,682 unique targets. This was performed with a 5$^{\prime\prime}$ crossmatch between the DESI target right ascension and declination of the pointing and the \textit{Gaia} coordinates. Since the main objective here was to search for inhomogeneous surface composition white dwarfs that switch between DA and DB, we only use data from the DESI blue arm to reduce confusion noise and improve sensitivity, as this arm contains the largest number of spectral lines. The spectra were normalised and dereddened in the same way as described in Section~\ref{subsec:methodsSpectra}.

\begin{figure}
    \centering
    \includegraphics[width=\columnwidth, clip, trim={0.25cm 0.25cm 0.25cm 0.25cm}]{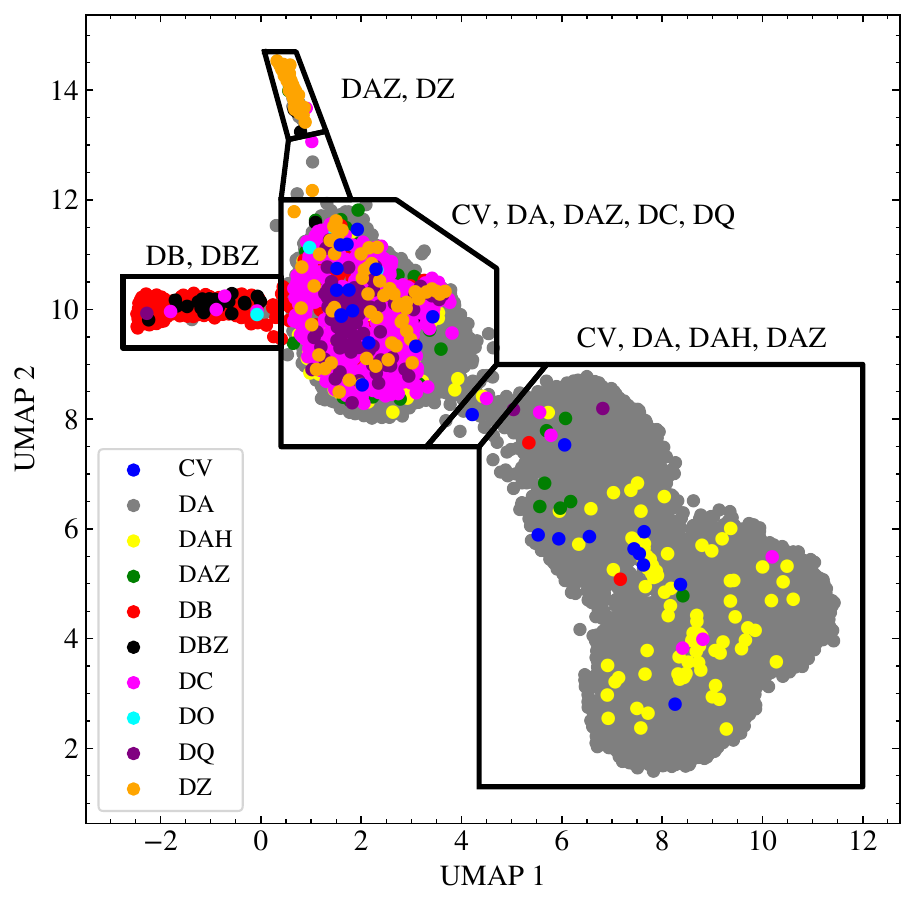}
    \caption{A UMAP representation from normalised spectra of the DESI blue arm. UMAP~1 and UMAP~2 in this figure are entirely independent to those labels in Fig.~\ref{fig:UMAPsingleClassifier}, since the input dataset is distinct. Objects shown here are the 21\,344 that passed all spectroscopic data cuts, and were used to identify class structure. Box labels describe the category of objects that fall within. Using these boxes, the full set of approximately 49\,682 unique DESI white dwarf candidates were investigated for drastic changes, being where two exposures fall in different boxes, by projection onto this UMAP coordinate space. The two boxes without labels are drawn to cover intermediate positions between the main boxes of interest, where an object could fall if it is transitioning between spectral types.}
    \label{fig:UMAPdoubleFaced}
\end{figure}

We first inserted the co-added spectra into a two-component UMAP routine with spectral types inferred from the single star classifier. We emphasise that this new UMAP coordinate projection is separate and unique from that in Fig.~\ref{fig:UMAPsingleClassifier}. The resultant UMAP coordinates were plotted as in Fig.~\ref{fig:UMAPdoubleFaced} and spectral types were used to categorise groupings. The new UMAP dimensionality reduction demonstrated a clear separation of classes, which is also labelled around boxes in Fig.~\ref{fig:UMAPdoubleFaced}. We then projected individual, single-epoch spectra onto the same UMAP visualisation and searched for cases where at least two spectra from a source fall in different boxes and inspected the results visually to search for drastic differences in spectral features. 

Multiple systems appeared due to two nearby, spatially resolved targets falling in different boxes, large signal-to-noise differences between spectra (as DA could be confused for a DC in a normalised spectrum at lower signals) or defects in the data products. Because of this and for the sake of time, we restricted the visual inspection to physically motivated cases where: 1) A target falls into the `DB+DBZ/DZB' box and any box where a DA could fall, for the inhomogeneous composition science case. 2) At least one spectrum falls in the DZ box, to probe variability in metal accretion. 3) At least one exposure falls in a box containing CVs, to search for mass transferring binaries that show a typical white dwarf spectrum and an outbursting state over the course of the full timespan of DESI DR1 data. Targets that were visually vetted and flagged as interesting are now described, where all were inspected for spatially resolved, nearby sources in \textit{Gaia} and removed if the case.

\begin{figure*}
    \centering
    \includegraphics[clip, trim={0cm 0.25cm 0.2cm 0.2cm}, width=\textwidth]{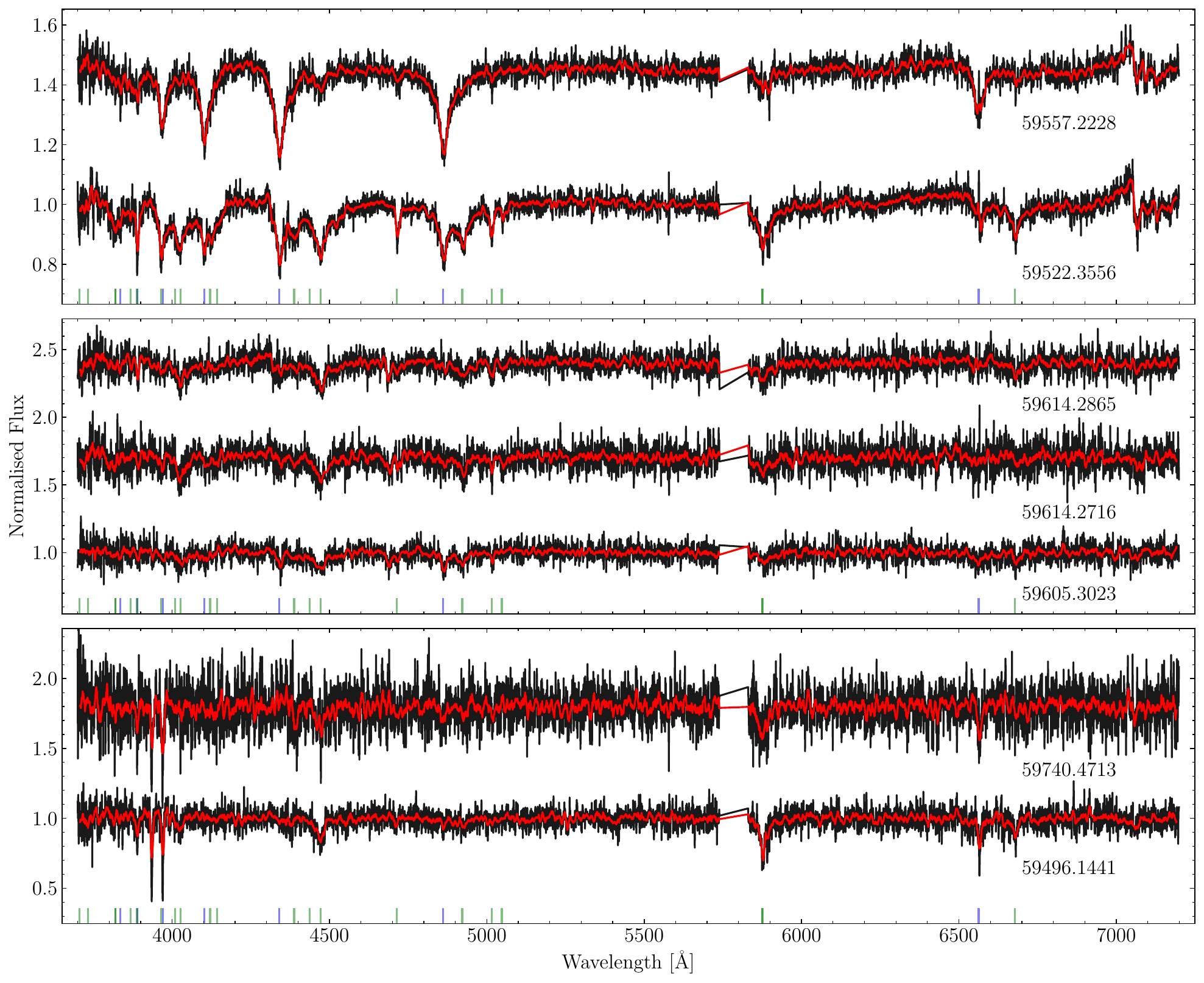}
    \caption{A figure displaying all normalised, single epoch spectra of the new, inhomogeneous surface composition (`double-faced') white dwarfs identified in Section~\ref{subsec:resultsSpectralTypeChanges}. \textit{Top:} WDJ022228.39+283007.72, \textit{middle:} WDJ091748.20+001041.72, \textit{bottom:} WDJ213146.85+025518.46. The reduced spectra are in black with a smoothed flux plotted red. The wavelengths corresponding to hydrogen and helium~I lines are vertically plotted just above the lower x-axes in blue and green, respectively. The mid-exposure modified BJD is written beneath each spectrum.  Blue arm data alone was input for UMAP analysis, while here the full visible spectrum is shown. Spectra are offset, each originally having a continuum normalisation equal to one.}
    \label{fig:WDJ022228.39+283007.72}
\end{figure*}

We successfully discovered three new candidate, inhomogeneous surface composition (`double-faced') white dwarfs in DESI with this new methodology. The first and perhaps most striking case is WDJ022228.39+283007.72 (KUV~02196+2816), with the two spectra in DESI plotted in Fig.~\ref{fig:WDJ022228.39+283007.72}. One spectrum is dominated by Balmer lines with a faint detection of 4471\AA~He~I absorption, while the other shows a mix of Balmer and He~I absorption features. This source was prematurely identified as a DA+DB double degenerate by analysis of a single spectrum \citep[][]{Limoges2009}, owing to the fact that neither a DA, DB or mixed hydrogen-helium atmosphere well fits the observations. Their theory could still align with the DESI observations for a near-perfectly edge-on binary and a chance eclipse at the time of observation, but the surface gravity identified by these authors for the DA and DB component is identical at 8.09\,dex. Although not impossible, this makes the chance of the helium-line component being eclipsed even less likely than it already was and, with the new DESI data, points in support of a single white dwarf with inhomogeneous surface composition. Additionally for this source, we see a main sequence companion in the near-infrared data.

\begin{figure}
    \centering
    \includegraphics[width=\columnwidth, trim={0.3cm 0.35cm 0.25cm 0.25cm}]{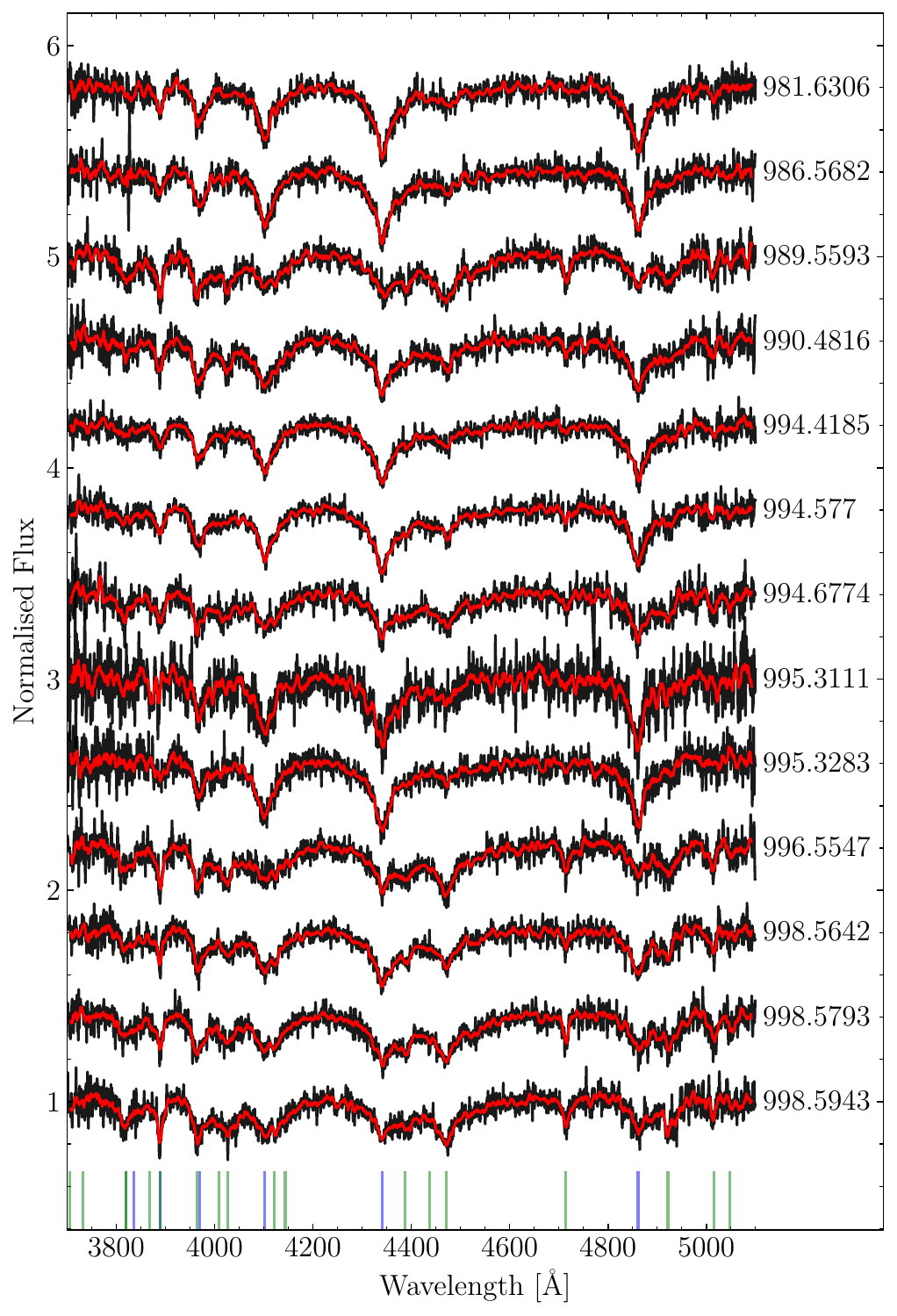}
    \caption{Normalised spectra of the inhomogeneous surface composition white dwarf WDJ022228.39+283007.72 that were obtained on the NOT. The blue and green vertical lines at the bottom correspond to the wavelengths of Balmer and He~I lines, respectively. The depth of He~I~4471\AA~changes across each exposures, while other He lines completely appear or disappear depending on phase. Photometry of the source combined with the spectral changes as a function of phase indicate the spin period to be 3.497\,hr, 3.051\,hr or 4.095\,hr, listed in decreasing Lomb Scargle periodogram powers. The mid-exposure Barycentric Julian Date minus 2\,460\,000 is stated on the right of each spectrum.}
    \label{fig:doublefacedNOTdata}
\end{figure}

The spectral change of this target is so vivid that we decided to obtain follow-up observations of the source and ascertain whether it is an inhomogeneous surface composition white dwarf. 1200\,s exposures (the same exposure time as the DESI data) were taken with the Alhambra Faint Object Spectrograph and Camera (ALFOSC) on the Nordic Optical Telescope using the g18 grism, giving a wavelength coverage of 3450--5350\AA. A 0.5$^{\prime\prime}$ slit width was employed for a spectral resolution of $R=\lambda/\Delta\lambda=2000$. All spectra were bias subtracted, flat fielded and flux calibrated with a spectrophotometric standard star using the \textsc{pypeit} package \citep[][]{pypeit}. A ThAr arc was taken before each exposure for wavelength calibration, and all reduced spectra of the target are plotted in Fig.~\ref{fig:doublefacedNOTdata}.

Spectra were spaced across multiple nights to better sample phase space, while three consecutive spectra were taken to be sensitive to short-term variability. Indeed, these three appear as near identical, implying that the spin period is at least a couple of hours. Having a variety of spectral line depths with blends of hydrogen Balmer lines and He~I lines clearly confirms that the source can not be an eclipsing white dwarf binary, as would be required to support the results of \citet[][]{Limoges2009}, and is instead an inhomogeneous surface composition white dwarf. Additionally, three spectra were taken on the same night ($\text{JD}=2460994$) spread across multiple hours to be better sensitive 2--10\,hr spin periods. Within these spectra, we see a transition from just Balmer lines with He~4471\AA~into a mixture of Balmer and other He~I lines, showing the gradual increase in the fraction of surface helium from the change in the line of sight.

Time-series photometry from the Zwicky Transient Facility \citep[ZTF,][]{ZTF} and the Asteroid Terrestrial-impact Last Alert System \citep[ATLAS,][]{ATLAS} was retrieved for WDJ022228.39+283007.72 to search for variability in its light curve. Data from individual filters were analysed with Lomb-Scargle periodograms \citep[][]{Lomb1976, Scargle1982} and a multi-band periodogram was also applied to all data combined \citep[][]{VanderPlas2015}, for which data from each filter is modelled separately but with a consistent phasing. The best-fit periods are consistent across each dataset, showing an approximately 5\% percentage flux amplitude. The five highest power peaks were manually validated by phase-folding the NOT spectra timestamps and ensuring that two spectra taken at the same phase look identical, excluding the DESI spectra since they were observed over three years earlier and could suffer from small cumulative errors in the best-fit periods. Doing so, we are able to restrict the spin period to 3 solutions, being 3.497\,hr, 3.051\,hr and 4.095\,hr, listed in decreasing power. This verified discovery and result alone shows that dimensional reduction techniques such as UMAP have a strong application in the analysis of time-series spectroscopy from MOS surveys.

The other two candidate inhomogeneous surface composition white dwarfs never had spectroscopic results presented in the literature to the best of our knowledge. The second source is WDJ213146.85+025518.46, which has a varying He~I absorption strength between the two available exposures (most visible at 4471\AA~and 6678\AA) and in only one spectrum shows faint H$\beta$ and H$\gamma$ absorption features. Additionally, a Ca~II doublet appears with a varying relative strength depending on the visible atmospheric composition. The best-fit temperatures to \textit{Gaia} photometry are $T_{\textrm{eff, He}}=13\,600\pm1900$\,K for a hydrogen-poor atmosphere and  $T_{\textrm{eff, H}}=14\,700\pm1600$\,K \citep[][]{NicolaGaia2021} for a hydrogen-rich one. One spectrum has a significantly lower signal-to-noise ratio than the other, so this detection could be spurious, but if the candidate is confirmed it would likely be the coolest inhomogeneous hydrogen-helium surface composition white dwarf to date after J0847+4842 \citep[][]{Moss2025}. The final object is WDJ091748.20+001041.72, with the most obvious changing spectral features over time being the relative depths of 4471\AA~He~I, 6678\AA~He~I and 4862\AA~H$\beta$. \citet[][]{NicolaGaia2021} find an effective temperature of $T_\textrm{eff, H}=34\,600\pm6\,500$\,K, placing this star in a far earlier white dwarf evolutionary phase and being similar to the `double-faced' white dwarf ZTF~J203349.8+322901.1 \citep[][]{Caiazzo2023}. All DESI spectra of these other two inhomogeneous surface composition objects are also plotted in Fig.~\ref{fig:WDJ022228.39+283007.72}. 

We emphasise that our UMAP technique is sensitive to the most apparent of changes between multiple spectra of the same source. Hence, it is probable that slightly spectrally changing inhomogeneous surface composition white dwarfs have passed us by, undetected in the DESI DR1 footprint. Many objects currently have a single spectrum in DESI as well, bringing hope to the detection of many more of these curious objects in the near future. A similar approach could be taken to SDSS or other MOS survey spectra for targets with consistently good signal-to-noise in multiple exposure or unique night data, which we promote for future work.

\begin{figure*}
    \centering
    \includegraphics[width=\textwidth, trim={0.25cm 0.35cm 0.3cm 0.3cm}]{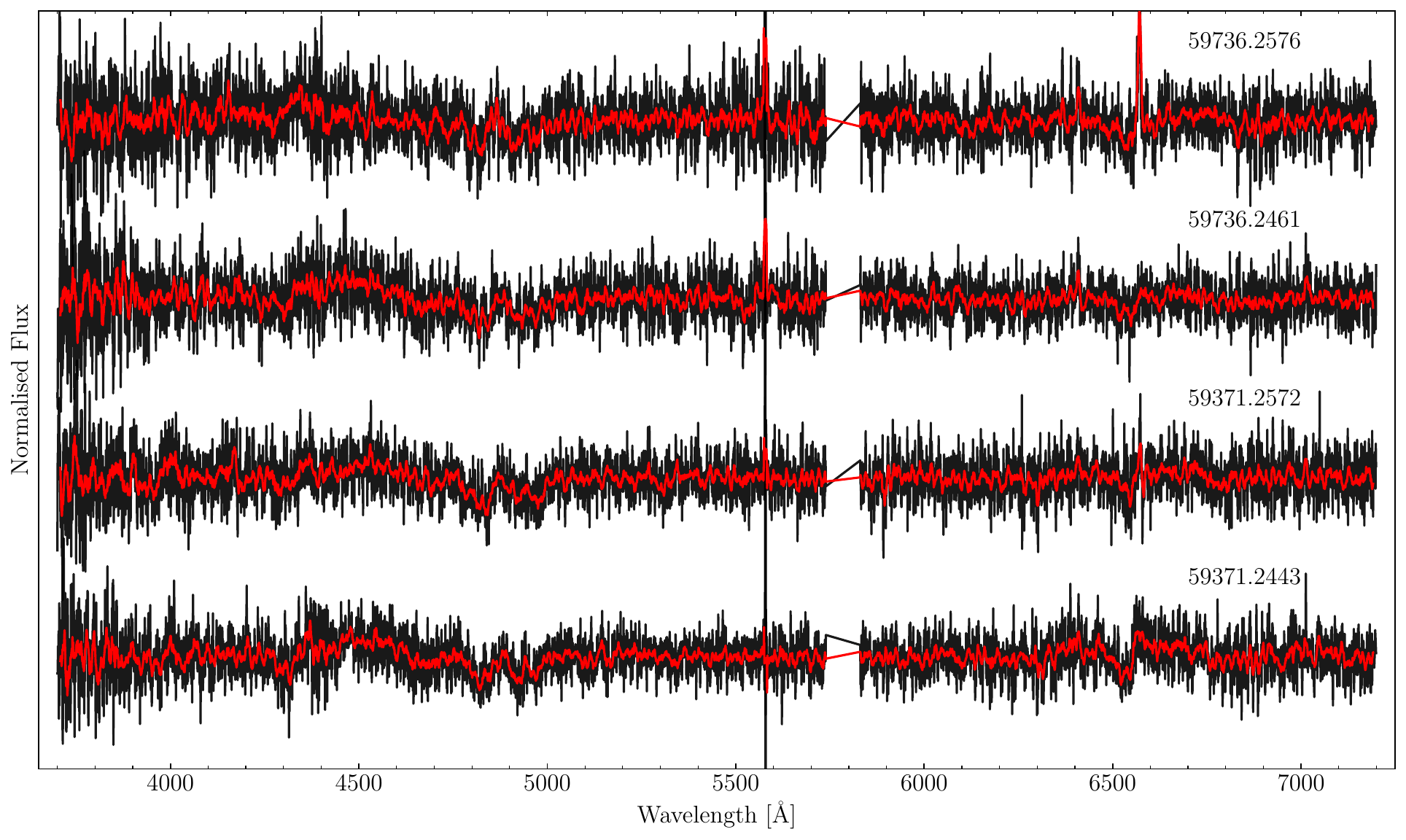}
    \caption{Individual spectra of WDJ150218.87+023054.98 -- an object classed as a DAH white dwarf by our single star classifier, which shows a varying Zeeman splitting as a function of a rotational phase and a strong H$\alpha$ emission line variability. The differences were drastic enough for the object to fall into different boxes for different exposures in Fig.~\ref{fig:UMAPdoubleFaced}. The mid-exposure Modified Barycentric Julian Date is stated above each spectrum.}
    \label{fig:WDJ125715.54+341439.53}
\end{figure*}

Our search retrieved other objects of clearly changing spectroscopic signatures. WDJ150218.87+023054.98 is a magnetic white dwarf with a strongly varying magnetic field strength dependent on the viewing angle as the white dwarf rotates (Fig.~\ref{fig:WDJ125715.54+341439.53}). Strong spectral differences are especially visible below 4500\AA, which are sufficient to cause the object to fall in two unique boxes in Fig.~\ref{fig:UMAPdoubleFaced}. The changes in spectral features across 20\,min exposures imply that the rotational period should be on the magnitude of hours. A similar signature is witnessed for WDJ125715.54+341439.53 which shows Zeeman split Balmer lines with a differing apparent field strength between two exposures split 100\,d apart, again strong enough to fall into two unique boxes. Lastly, WDJ131900.57+073752.42 exhibits He~I absorption features and an emission line at approximately 4300\AA, with 6680\AA~He~I absorption depth variability across the three unique nights of exposures.

\section{Flagging binarity with machine learning}
As previously mentioned, the combination of spectroscopy and photometry is not only crucial for single star fitting, but also revealing the presence of a stellar companion. DESI comes with a higher spectral sampling and resolution compared to SDSS, such that its data products should be more sensitive to small changes in the broadness of spectral lines too. To that extent, we investigated the capability of machine learning techniques in the detection of binary star systems with DESI spectra. The motivation for this exploration is to see if one can immediately remove certain objects from a sample of white dwarfs that could contaminate population statistics, for example the white dwarf mass distribution.

As a training set, all objects have the requirement of showing Balmer absorption lines, since hydrogen lines are almost always present for every object in the full sample of known double white dwarfs. We use all single and double white dwarfs given in \citet[][]{Munday2024DBL}, the most up-to-date list of compact double white dwarf binaries found at \textsc{https://github.com/JamesMunday98/CloseDWDbinaries}, all single and double white dwarfs analysed in \citet[][]{Napiwotzki2020spy} and finally all wide, double white dwarf binaries found in the catalogue of \citet[][]{ElBadry2021edr3properMotion} which have an angular separation of less than 1$^{\prime\prime}$. A maximum of 1$^{\prime\prime}$ was chosen as the fiber width of DESI is 1.5$^{\prime\prime}$, meaning that, regardless of which star's sky coordinates are selected as the fiber positioning centre, the vast majority of flux from both stars will enter the fiber for typical site seeing conditions. The \citet{ElBadry2021edr3properMotion} objects that had DESI spectra were visually inspected to ensure that all showed Balmer absorption lines.

Noteworthy, some of the single star objects in this training set may actually be binary systems where the companion contributes a significant amount of light. This will not matter for the \citet[][]{ElBadry2021edr3properMotion} objects since the stars are resolved and this is unlikely to be problematic for the \citet[][]{Munday2024DBL} objects since fitting of simultaneous spectroscopic and photometric data occurred. However, this was not the case in \citet[][]{Napiwotzki2020spy}, which could pollute our sample, but the inclusion of these objects is beneficial for augmenting the training set sample size with few suspected impurities. Furthermore, our selection cut of a $P_\textrm{WD}>0.5$ from the \citet[][]{NicolaGaia2021} catalogue removes many double white dwarfs which host an extremely low mass white dwarf \citep[e.g.][]{Brown2020elmNorthFinal, Kosakowski2023elmSouth}. But, in these cases, the extremely low mass white dwarf almost completely outshines the companion and binarity is only recognisable through radial velocity or photometric variability. Hence, omission of these objects is positive for this investigation which depends on a single, co-added spectrum. Double-lined (spectral signatures from both stars are recognisable) or single-lined (spectral signatures from just one star) double white dwarfs were combined into a generic category encompassing all binary sources. This was needed since DESI spectra taken on many nights for the same target are co-added, which would lead to a smeared double-lined feature, or because unlucky orbital phase sampling could hide any double-lined detection.

We employed a neural network that is similar to that described in Section~\ref{sec:neuralNetwork}, but, because a source is single or binary, we only have two unique classes. A binary cross-entropy loss function is more appropriate, which reports a single value for each source that evaluates how likely a binary star system is. This differs from the outputs of the neural network in Section~\ref{sec:neuralNetwork} as here a full matrix of probabilities for each class was reported. The single star and binary star classes were equally weighted in the  model. Finally, because a larger wavelength coverage is more important for uncovering a companion star, we included SDSS $u$ photometry (or synthetic SDSS $u$ photometry from \textit{Gaia}~XP spectra) when the error in magnitude was below 0.3\,mag.

Since we required that targets must show Balmer lines, we normalised the co-added spectra in the same way as Section~\ref{subsec:methodsSpectra} but only clipping the Balmer line wavelength regions listed in Appendix~\ref{Appendix:masking}. As a last adaptation for the task, we removed all spectroscopic data below 3860\AA, which generally has a lower signal-to-noise ratio and weaker spectral line signatures. Overall, 224 objects out of 799 from our high-confidence binary/single star input catalogue appear in DESI DR1, with 170 objects passing signal-to-noise and photometric quality cuts (Section~\ref{sec:Methods}).

\begin{figure*}
    \centering
    \includegraphics[width=\textwidth, clip, trim={0.5cm 0.6cm 0.5cm 0.5cm}]{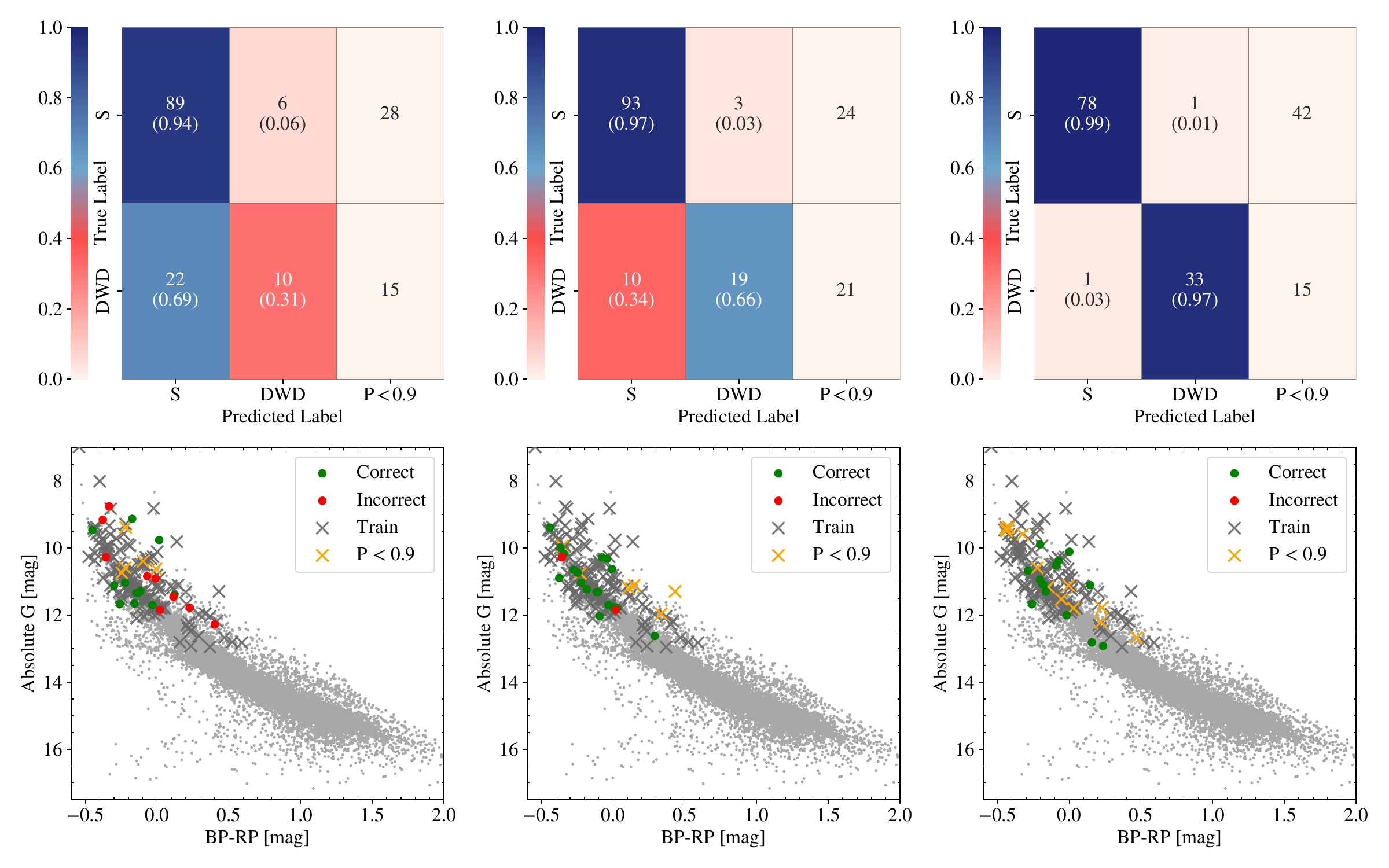}
    \caption{Performance of our single star (S) or double white dwarf (DWD) classifier. The top row of plots shows K-fold confusion matrices for trained networks based on spectral information only (left), a 50\%-50\% weighting between the DESI spectroscopy and Pan-STARRS photometry (middle), and Pan-STARRS photometry only (right). In the bottom plot are results for one of the five 80\%-20\% train-test iterations in the binary classifier. Grey crosses are the training set and the circular dots are the tested sample. Correctly classified points are in green, incorrectly classified points are in red. The background points are all white dwarfs in \textit{Gaia} within 100\,pc that have $P_\textrm{WD}>0.5$ \citep[][]{NicolaGaia2021}. The \textit{Gaia} Hertzsprung-Russell diagram is plotted to ease human interpretation, while extra absolute magnitude information was supplied to the machine for the $u,g,r,i,z,y$ bands.}
    \label{fig:binaryClassifierResultsSpecHybridPhot}
\end{figure*}

Correlation matrices for spectroscopy to photometry weightings of 100\%--0\%, 50\%--50\% and 0\%--100\% are shown in Fig.~\ref{fig:binaryClassifierResultsSpecHybridPhot}. We chose a stricter probability threshold when considering a star as single or binary of $P>0.9$ in an attempt remove confusion noise which could stem from e.g. all single-lined sources being identified as a single white dwarf or all over-luminous sources deemed as a binary. We see that binary candidates are identifiable using just the spectroscopy, photometry or a combination of the two, since the false positive rates of a single star being identified as a double white dwarf is close to zero. The sample of double white dwarfs is relatively small, yet, interestingly, clear evidence is shown that using a 100\% weighting of the $u,g,r,i,z,y$ absolute photometry is better than including spectroscopy at all.

It is clear that larger sample sizes with future MOS survey data releases would be able to give a more rounded training set while giving better statistics on the accuracy of machine learning algorithms to detect white dwarf binaries. Future efforts to discover more wide and compact binaries will be essential too, especially for fainter absolute magnitudes ($G\gtrapprox13$\,mag), which are not part of the input dataset. Though, in the present, our investigation shows a clear means to assist in efficiently removing binary impurity from large-scale samples using machine learning techniques when dealing with white dwarf spectra showing Balmer absorption lines.

\section{Conclusions}
We used machine learning to classify white dwarf spectral types, focussing on DESI DR1 as the best quality, large spectroscopic sample to date with a consistent data reduction, combined with absolute Pan-STARRS photometry. The neural network was trained and tested on the data products of the surveys, without the need for any white dwarf model atmospheric models, and is specialised to find the most common white dwarf spectral types, the presence of metals and the presence of magnetism. The DA and DB classes have a near 100\% classification accuracy, while the other primary spectral type classes have accuracies ranging from 85--95\%. Spectral types that indicate metals have an above 90\% accuracy, and are only confused when the metal line signature is very weak or when there are uncommon metal lines in the trained dataset.

We then demonstrated how the UMAP technique can be well employed to the data to visualise spectral class structure through dimensionality reduction. Distinct populations in UMAP coordinate space are revealed for DO/DOA/DAO white dwarfs, intermediate field strength DAH white dwarfs, CVs, DQs with clear Swan bands and metal polluted white dwarfs. We also showed how other specific populations, like DAHe white dwarfs, can be identified. This is an excellent technique to help correct catalogued spectral type inaccuracies. At the same time, it exemplifies how one can take combined spectroscopic and photometric measurements of white dwarfs, handle the data in a uniform way and semi-automatically isolate certain sub-populations of interest through UMAP coordinate space alone. Continuing with UMAP analyses, we searched DESI DR1 for situations where two individual exposures of the same target vary significantly, revealing three new inhomogeneous surface composition (`double-faced') white dwarfs along with a couple of objects with changing magnetism signatures. The nature of one of the inhomogeneous surface composition white dwarf detections was verified with phase-resolved time series spectra, indicating this technique as a powerful tool to quickly identify many similar systems in MOS surveys that have multiple spectroscopic epochs.

Finally, we investigated whether machine learning tools can be used to effectively identify double white dwarf binary star systems hiding amongst large white dwarf datasets, which would be ideal to eliminate when performing bulk single-star population statistics. DESI spectroscopy, Pan-STARRS photometry or a combination of each are able to recover white dwarf binaries candidates well, with a very small false positive rate of a single star being characterised as a binary. There is plenty of improvement that can be made to increase performance statistics, requiring larger training sample sizes especially for fainter absolute magnitudes, but our results demonstrate an effective means to efficiently isolate binaries in a single star dominated sample.

Our study clearly shows how machine learning techniques, or data visualisation techniques that are commonly used in the context of machine learning, are excellent complements to human classification in the big data era of astronomy. While they may confuse the most unique of objects with other, similar spectral types, which may be clearer to a human expert, the techniques demonstrated in this paper have the power to automate the classification of thousands of white dwarfs in a highly accurate manner in seconds. All methods outlined in this study can be applied to any MOS survey of white dwarfs, which will be vital for future DESI data releases, as well 4MOST \citep[][]{4most2019} and WEAVE \citep[][]{WEAVEwhitepaper2024} that should have a similar data quality. Batch population analyses are ideal to depend on machine learning as hundreds of thousands of white dwarf spectra arrive in the coming decade, often viewing the white dwarfs for the first time spectroscopically.

\section*{Acknowledgements}
This project has received funding from the European Research Council under the European Union’s Horizon 2020 research and innovation programme (Grant agreement numbers 101002408 – MOS100PC). IP acknowledges support from a Royal Society University Research Fellowship (URF\textbackslash R1\textbackslash 231496). 
TLK acknowledges support from a Warwick Astrophysics prize post-doctoral fellowship made possible thanks to a generous philanthropic donation. 

DJ acknowledges support from the Agencia Estatal de Investigaci\'on del Ministerio de Ciencia, Innovaci\'on y Universidades (MCIU/AEI) under grant ``Nebulosas planetarias como clave para comprender la evoluci\'on de estrellas binarias'' and the European Regional Development Fund (ERDF) with reference PID2022-136653NA-I00 (DOI:10.13039/501100011033). DJ and PS acknowledge support from the Agencia Estatal de Investigaci\'on del Ministerio de Ciencia, Innovaci\'on y Universidades (MCIU/AEI) under grant ``Revolucionando el conocimiento de la evoluci\'on de estrellas poco masivas'' and the European Union NextGenerationEU/PRTR with reference CNS2023-143910 (DOI:10.13039/501100011033).

Based on observations made with the Nordic Optical Telescope, owned in collaboration by the University of Turku and Aarhus University, and operated jointly by Aarhus University, the University of Turku and the University of Oslo, representing Denmark, Finland and Norway, the University of Iceland and Stockholm University at the Observatorio del Roque de los Muchachos, La Palma, Spain, of the Instituto de Astrof\'isica de Canarias.  The NOT data presented here were obtained, under program IDs P72-204 (70-NOT7/25B) and P72-401, with ALFOSC which is provided by the Instituto de Astrofisica de Andalucia (IAA) under a joint agreement with the University of Copenhagen and NOT.

%%%%%%%%%%%%%%%%%%%%%%%%%%%%%%%%%%%%%%%%%%%%%%%%%%
\section*{Data Availability}
All DESI DR1 co-added or single epoch spectra are downloadable through the mission's public archive. Our training sample with classifications is available at \textsc{https://github.com/JamesMunday98/MachineLearningWD}. The machine learning spectral types deduced when applied to the 21\,344 DESI DR1 sample which pass our data quality cuts will be available on the same site once the principal investigator of the DESI white dwarf team has released theirs, to enable a direct comparison between human and machine classifications. Reduced NOT spectra of the inhomogeneous surface composition white dwarf WDJ022228.39+283007.72 will be made available upon request to the authors.

%%%%%%%%%%%%%%%%%%%% REFERENCES %%%%%%%%%%%%%%%%%%

% The best way to enter references is to use BibTeX:

\bibliographystyle{mnras}
\bibliography{mnras_template} % if your bibtex file is called example.bib

% Alternatively you could enter them by hand, like this:
% This method is tedious and prone to error if you have lots of references
%\begin{thebibliography}{99}
%\bibitem[\protect\citeauthoryear{Author}{2012}]{Author2012}
%Author A.~N., 2013, Journal of Improbable Astronomy, 1, 1
%\bibitem[\protect\citeauthoryear{Others}{2013}]{Others2013}
%Others S., 2012, Journal of Interesting Stuff, 17, 198
%\end{thebibliography}

%%%%%%%%%%%%%%%%%%%%%%%%%%%%%%%%%%%%%%%%%%%%%%%%%%

%%%%%%%%%%%%%%%%% APPENDICES %%%%%%%%%%%%%%%%%%%%%

\appendix

\section{Masked spectral regions in continuum normalisation}
\label{Appendix:masking}
Spectral regions that commonly show spectral lines were masked in an attempt to maintain a continuum normalisation equal to one. A trade off between maintaining spectral coverage had to be performed at times to maintain a good normalisation for the majority of spectra. For instance, DQ white dwarfs with broad and deep Swan bands would require a masking that would prove detrimental to the normalisation of DA and DB spectral types. However, the common method in normalising objects of all spectral types with an identical approach means that artefacts in non-perfect normalisation are consistent between the training and the test datasets, allowing all to be recognised by our neural network. 

The clipped wavelengths regions were as follows:
\begin{itemize}
    \item Balmer lines: 6412--6712\AA, 4742--4962\AA, 4242--4442\AA, 4050--4187\AA, 3930--4025\AA, 3865--3920\AA, 3815--3855\AA
    \item Helium I lines: 5825--5925\AA, 5000--5030\AA, 5042--5060\AA, 4705--4730\AA, 4320--4580\AA, 3990--4073\AA, 3788--3850\AA, 5568--5605\AA, 7035--7105\AA, 7260--7315\AA
    \item Metal lines: 5140--5200\AA, 5795--5830\AA, 4211--4245\AA
    \item DQ: 5420--5450\AA, 6290--6340\AA, 5990--6080\AA, 4620--4670\AA
\end{itemize}

While they can be either narrow or broad, calcium lines were intentionally not masked to allow for a better blue normalisation in the other spectral classes, but we tackled this problem by rejecting binned points that lie at least 2.5$\sigma$ below the residual of the continuum fit (Section~\ref{subsec:methodsSpectra}).

%%%%%%%%%%%%%%%%%%%%%%%%%%%%%%%%%%%%%%%%%%%%%%%%%%

% Don't change these lines
\bsp	% typesetting comment
\label{lastpage}
\end{document}